\documentclass[letterpaper]{article} 
\usepackage{aaai24}  
\usepackage{times}  
\usepackage{helvet}  
\usepackage{courier}  
\usepackage[hyphens]{url}  
\usepackage{graphicx} 
\usepackage{natbib}  
\usepackage{caption} 
\usepackage{booktabs}
\usepackage{algorithm}
\usepackage{algorithmic}

\usepackage{newfloat}
\usepackage{listings}
\DeclareCaptionStyle{ruled}{labelfont=normalfont,labelsep=colon,strut=off} 
\floatstyle{ruled}
\newfloat{listing}{tb}{lst}{}
\floatname{listing}{Listing}
\usepackage{algorithm}
\usepackage{algorithmic}
\usepackage{multirow}
\usepackage{amsmath} 
\usepackage{enumitem}

\usepackage{caption} 

\usepackage{etoolbox} 

\AtBeginEnvironment{tabular}{\small}

\title{Wikiformer: Pre-training with Structured Information of Wikipedia for Ad-hoc Retrieval}
\author {
    Weihang Su\textsuperscript{\rm 1}\thanks{First Author: swh22@mails.tsinghua.edu.cn},
    Qingyao Ai\textsuperscript{\rm 2}\thanks{Corresponding Author: aiqy@tsinghua.edu.cn},
    Xiangsheng Li\textsuperscript{\rm 2},
    Jia Chen\textsuperscript{\rm 2},
    Yiqun Liu\textsuperscript{\rm 2},
    Xiaolong Wu\textsuperscript{\rm 3},
    Shengluan Hou\textsuperscript{\rm 3}
}
\affiliations {
        \textsuperscript{\rm 1}Quan Cheng Laboratory \&
DCST, Tsinghua University \&
Zhongguancun Laboratory,
Beijing, China\\
    \textsuperscript{\rm 2}DCST, Tsinghua University \&
Zhongguancun Laboratory,
Beijing, China\\
    \textsuperscript{\rm 3}Huawei Poisson Lab\\
}

\usepackage{bibentry}
\usepackage{etoolbox}
\AtBeginEnvironment{tabular}{\small}
\AtBeginEnvironment{table}{\small}
\begin{document}

\maketitle

\begin{abstract}

With the development of deep learning and natural language processing techniques, pre-trained language models have been widely used to solve information retrieval (IR) problems. Benefiting from the pre-training and fine-tuning paradigm, these models achieve state-of-the-art performance. In previous works, plain texts in Wikipedia have been widely used in the pre-training stage. However, the rich structured information in Wikipedia, such as the titles, abstracts, hierarchical heading (multi-level title) structure, relationship between articles, references, hyperlink structures, and the writing organizations, has not been fully explored. In this paper, we devise four pre-training objectives tailored for IR tasks based on the structured knowledge of Wikipedia. Compared to existing pre-training methods, our approach can better capture the semantic knowledge in the training corpus by leveraging the human-edited structured data from Wikipedia. Experimental results on multiple IR benchmark datasets show the superior performance of our model in both zero-shot and fine-tuning settings compared to existing strong retrieval baselines. Besides, experimental results in biomedical and legal domains demonstrate that our approach achieves better performance in vertical domains compared to previous models, especially in scenarios where long text similarity matching is needed. The code is available at {https://github.com/oneal2000/Wikiformer}.
\end{abstract}

\section{Introduction}

Pre-trained Language Models (PLMs) have achieved great success in the field of Natural Language Processing (NLP)\cite{devlin2018bert, vaswani2017attention, yang2019xlnet, liu2019roberta, yasunaga2022linkbert}. These models are firstly pre-trained on a large-scale unlabeled text corpus and then fine-tuned on certain downstream tasks. The pre-training and fine-tuning paradigm have achieved state-of-the-art performance in many downstream NLP tasks. Recently, it has also attracted the attention of the Information Retrieval (IR) community. Besides directly applying PLMs to solve downstream IR tasks\cite{nogueira2019passage}, IR researchers have also developed several pre-training methods tailored for IR tasks, especially ad-hoc search\cite{ma2021prop, ma2021pre, ma2021b, chang2020pre}. These studies have shown promising results in conducting IR-specific pre-trained models for downstream tasks.

\begin{figure}[h!]
\centering
    \includegraphics[width=\columnwidth]{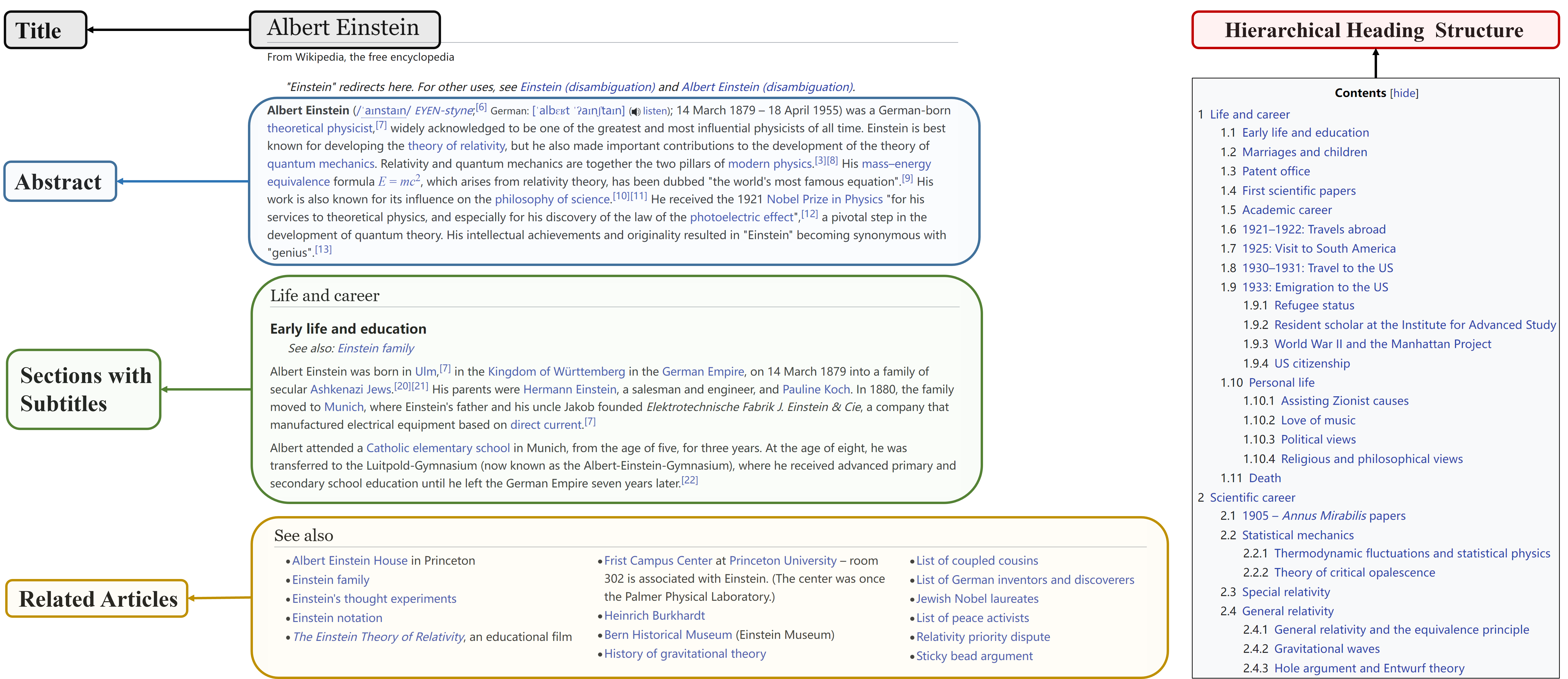}
    \caption{Rich structured information of Wikipedia.}
    \label{wiki}
\end{figure}

As one of the largest online knowledge bases, Wikipedia has been widely used as the pre-training corpus. In previous works, IR researchers have devised several pre-training tasks by leveraging the rich textual contents in Wikipedia. For example, PROP\cite{ma2021prop} utilizes pure texts in Wikipedia, while HARP\cite{ma2021pre} utilizes hyperlinks and anchor texts in the web pages. However, as shown in Figure \ref{wiki}, there's more rich knowledge brought by the structured information of Wikipedia, which, to the best of our knowledge, has not been exploited in existing studies. For example, the abstract section of Wikipedia is the summarization of an article. When the user’s query is the title of an article, the abstract section is more likely to match the user’s information needs compared to other sections within the same article. In addition, every article on Wikipedia has a hierarchical heading (multi-level title) structure, the subtitle is always the representative words or summarization of the corresponding section. Besides, different subsections of the same section share similar ideas. The relationship between different articles also contains rich information, e.g.,  the See Also section links one article to other articles that contain additional or similar information. Whether this structured knowledge could benefit the pre-trained models for IR remains mostly unknown.

To better incorporate the knowledge of Wikipedia into the pre-training stage, we propose a framework named Wikiformer that fully utilizes the structured information of Wikipedia in the pre-training stage. Wikiformer mainly includes four pre-training tasks: 1) Simulated Re-ranking (SRR), 2) Representative Words Identification (RWI), 3) Abstract Texts Identification (ATI), and 4) Long Texts Matching (LTM). These tasks use the title, subtitles, abstract, hyperlinks, and heading hierarchies to construct pseudo query-document pairs for the pre-training of the retrieval model. Each of them captures the needs of retrieval and ranking in different granularities from different angles. To evaluate the effectiveness of the above pre-training tasks, we test the performance of our model on several IR benchmarks in zero-shot and fine-tuning settings. In the zero-shot setting, no supervised data is used for fine-tuning. Since the fine-tuning process gradually updates the parameters of PLMs, zero-shot performance is a more direct metric to evaluate the effectiveness of pre-training methods. The experimental results show that Wikiformer can significantly outperform traditional methods, state-of-the-art neural ranking models, and existing pre-trained models for IR in multiple domains with or without human-annotated data. 

In summary, the contributions of our work are three folds:

\begin{itemize}
\item We propose a novel pre-training framework, i.e., Wikiformer,  that makes full use of the structured knowledge of Wikipedia. 
\item We propose four learning objectives based on pseudo query-document pair sampling during pre-training.  Tailored for IR tasks such as retrieval and document re-ranking, these objectives can better help the model analyze the relevance between queries and documents.
\item We evaluate Wikiformer on multiple IR benchmark datasets, and the experimental results show that Wikiformer outperforms state-of-the-art methods in both zero-shot and fine-tuning settings in multiple domains. 
\end{itemize}

\section{Related Work}
\subsection{Pre-trained Language Models}
Pre-trained Language Models (PLMs) have achieved great success in recent years\cite{devlin2018bert, vaswani2017attention, yang2019xlnet, liu2019roberta, yasunaga2022linkbert}. These models are firstly trained on large-scale unlabeled text corpora and then fine-tuned on certain downstream tasks with labeled data. Benefiting from the self-supervised learning on a large-scale pre-training corpus, these models own a powerful ability on contextual text representation. 

Among these PLMs, Transformer based models\cite{vaswani2017attention} show great performance in most downstream NLP tasks. One of the remarkable examples is the BERT model\cite{devlin2018bert}, a bi-directional Transformer based pre-trained language model. BERT has two self-supervised tasks in the pre-training stage: Masked Language Modeling (MLM) and Next Sentence Prediction (NSP). Following BERT, researchers redesign and optimize the pre-training tasks of PLMs. For example, Roberta\cite{liu2019roberta} uses a dynamic masking strategy and is trained on a larger text corpus. In addition, some researchers explore the integration of structured information into PLMs\cite{yasunaga2022linkbert,colon2021combining,kaur2022lm,zhang2019ernie}. For example, LinkBERT\cite{yasunaga2022linkbert} replaces the NSP task of BERT with the Document Relation Prediction (DRP) task, which enables the model to learn cross-document knowledge from hyperlinks among web pages. ERNIE\cite{zhang2019ernie} utilizes both textual corpora and Knowledge Graphs to train an enhanced PLM.

\subsection{Pre-training Methods Tailored for IR}
Considering the great success that PTMs have achieved in NLP tasks, the IR community begins to apply PLMs to solve IR tasks~\cite{li2023thuir,li2023towards,chen2023thuir,li2023constructing,ye2023relevance}, and devise pre-training methods tailored for IR\cite{ma2023caseencoder,ma2021prop, ma2021pre, ma2021b, chang2020pre,fan2021pre,guo2022webformer,chen2022axiomatically,su2023caseformer,su2023thuir2,li2023sailer}. For example, HARP\cite{ma2021pre} utilizes hyperlinks and anchor texts in the pre-training stage. As the anchor texts are edited by humans, constructing pseudo query-document pairs from them may be more reliable than an algorithm. Webformer\cite{guo2022webformer} is a pre-trained language model based on large-scale web pages, HTML tags, and the DOM (Document Object Model) tree structures of web pages. \citeauthor{ma2021prop}\cite{ma2021prop} devised a self-supervised learning task Representative Words Prediction (ROP) based on the Query Likelihood model and train the Transformer encoder with a self-supervised contrastive learning strategy. From another angle, ARES\cite{chen2022axiomatically} propose several pre-training objectives based on Axiomatic Regularization. Experimental results on several IR benchmarks show that ARES, PROP, Webformer, and HARP perform significantly better than traditional methods such as BM25 after fine-tuning. Also, some researchers explore incorporating structure information for entity retrieval\cite{gerritse2020graph, nikolaev2020joint, chatterjee2022bert, gerritse2022entity}.

Different from the above approaches, we propose four new pre-training objectives using the titles, abstracts, hierarchical heading (multi-level title) structure, relationship between articles, references, hyperlink structures, and the writing organizations of Wikipedia to leverage the wisdom of crowds brought by Wikipedia editors. Compared to previous work, Wikiformer captures more internal relationships between the paragraph structure in Wikipedia web pages, which helps it better model relevance matching.

\section{Methodology}
\label{section:method}
The main objective of our pre-training method is to leverage the structured knowledge and writing organization of Wikipedia for designing better pre-training tasks tailored for information retrieval. To achieve this, we propose four pre-training tasks based on the titles, abstracts, hierarchical heading (multi-level title) structure, the relationship between articles, and the writing organizations of Wikipedia.

In this section, we introduce the details of the pre-training tasks of our proposed model Wikiformer, including Simulated Re-ranking (SRR), Representative Words Identification (RWI), Abstract Texts Identification (ATI), and Long Texts Matching (LTM) tasks. 

\subsection{Simulated Re-ranking  (SRR)}
\label{s31}

\begin{figure}[h]
\centering
    \includegraphics[width=\columnwidth]{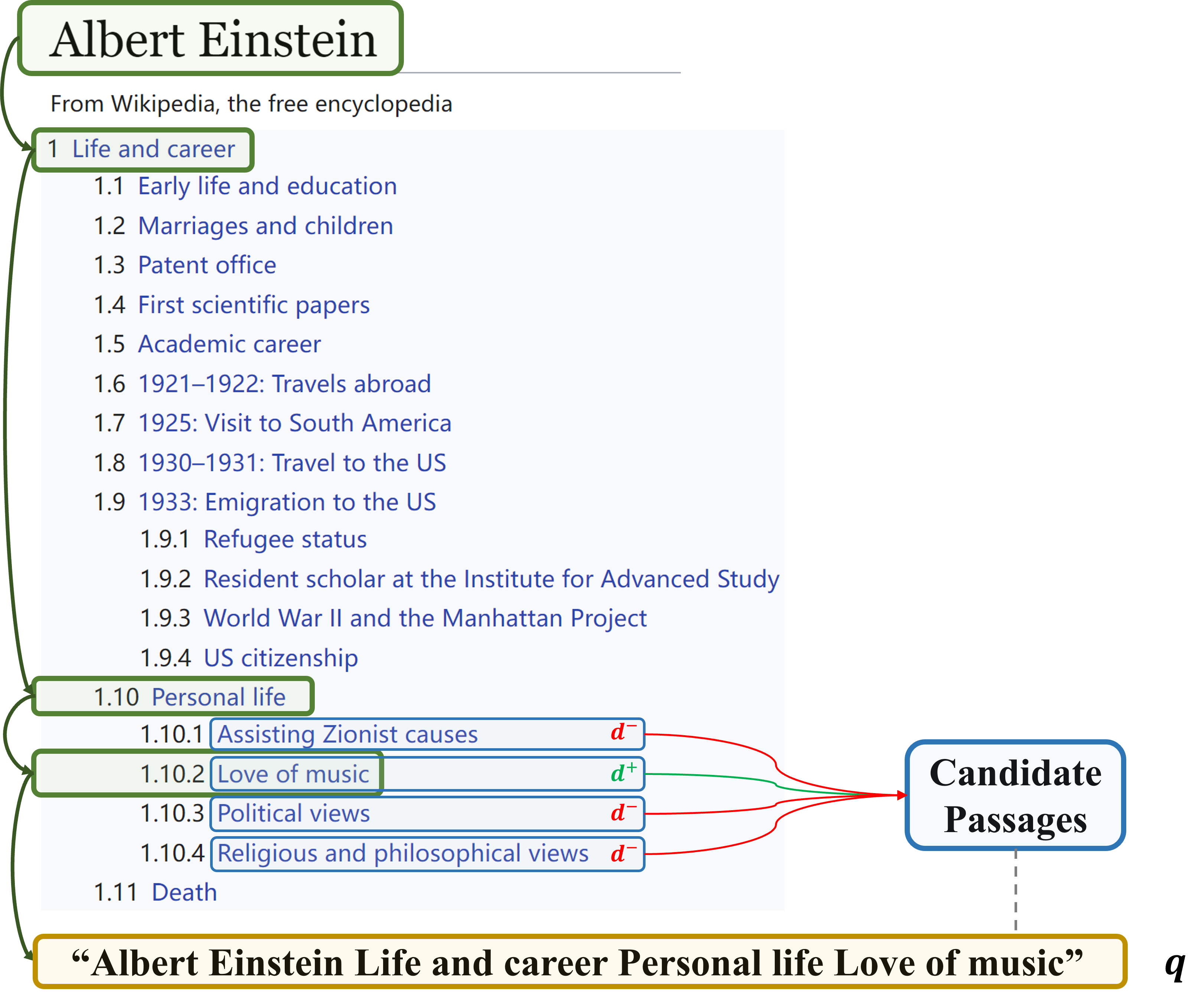}
    \caption{Pseudo query-document pairs generated from the tree structure of a Wikipedia article, where $q$ is the query, $d^+$ is the positive document and $d^-$ are negative documents.}
    \label{fig:sep_query}
\end{figure}

The SRR task is inspired by an important IR problem: document re-ranking. In general, the goal of the document re-ranking task is to sort a series of documents that are highly related to the query, and then select the ones that are most related to the query. According to the characteristics of this task, we aim to design a self-supervised learning task to select the most relevant document from a series of documents with similar contents. In the SRR task, we make full use of the hierarchical heading (multi-level title) structure of Wikipedia to achieve the above objective. Every article on Wikipedia is organized by the hierarchical heading (multi-level title) structure, the subtitle corresponding to a certain section tends to be the representative words or summarization of the text. Besides, different subsections of the same section share similar semantics. As a result, through this structure, we can obtain a series of texts that are highly similar but slightly different in content and generate the query through the multi-level titles as shown in Figure \ref{fig:sep_query}.

To be specific, we modeled each Wikipedia article into a tree structure namely Wiki Structure Tree (WST)  based on the hierarchical heading structure. It can be defined as: 

\vspace{1mm}
$ WST = <D, R>$, where $D$ is a finite set containing $n$ nodes, and $R$ is the root node of $ WST$. Each node in $D$ consists of two parts: the subtitle and its corresponding content. The root node $R$ contains the main title and the abstract of this article. Starting from the root node $R$, recursively take all the corresponding lower-level sections as its child nodes until every section in this article is added to the $WST$. 
\vspace{1mm}

After building $WST$, we use a contrastive sampling strategy to construct pseudo query-document pairs based on the tree. For a non-leaf node $F$ in the $WST$, we add all its child nodes to the set $S$. A node $d_i$ is randomly selected from $S$. Traversing from the root node to node $d_i$, all the titles on the path are put together to form a query $q$. This process is shown in Figure \ref{fig:sep_query}. The content of the node $d_i$ is defined as $d^+$, and the content of the other nodes in $S$ is defined as $d^-$. We use a Transformer based PLM to compute the relevance score of a pseudo query-document pair:
{\small
\begin{equation}
\label{equ_input}
Input = [CLS] query [SEP] document [SEP] 
\end{equation}

\begin{equation}
\label{equ_score}
score(query,document) = MLP(Transformer(Input) ) 
\end{equation}

}

\noindent where $Transformer(Input)$ is the vector representation of the "[CLS]" token. $MLP(\cdot)$ is a multi-layer perceptron that projects the [CLS] vector to a relevance score. For the loss function, we use the Softmax Cross Entropy Loss\cite{cao2007learning,ai2018learning,gao2021rethink} to optimize the Transformer based model, which is defined as:

\begin{equation}
    \label{eq:SRR}
   \mathcal{L}_{SRR} = -\log_{}{    \frac{exp(score(q,d^+))}{exp(score(q,d^+))+\sum_{d\in S} exp(score(q,d))}} 
\end{equation}

\noindent where $q$, and $d^+$ are defined above and $S$ is the set of all negative passages generated from $WST$.

\subsection{Representative Words Identification (RWI)}
\label{s32}
\begin{figure}[h]
\centering
    \includegraphics[width=\columnwidth]{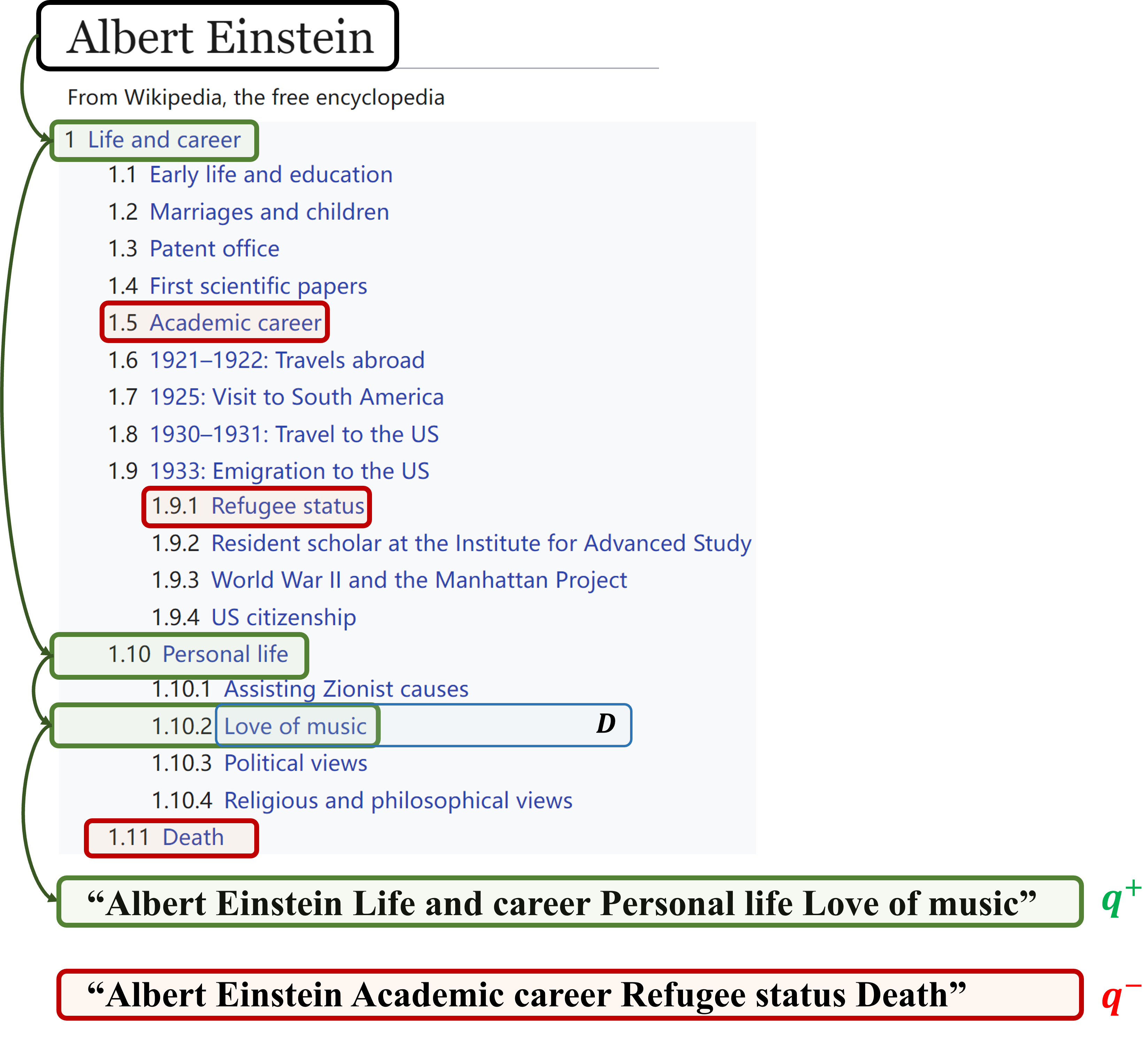}
    \caption{The contrastive sampling strategy of the RWI task, where $D$ is the document, $q+$ is the positive query and $q^-$ is the negative query.}
    \label{fig:TIP}
    
\end{figure}
  {
RWI task is inspired by an IR axiom which assumes that the user’s query is the representative words extracted from the relevant documents. According to the Wikipedia structure, we regard the subtitle of each section as representative words, and then we sample pseudo query-document pair via a simple strategy based on the hierarchical heading (multi-level title) structure, as shown in Figure \ref{fig:TIP}.

Specifically, pseudo query-document pairs are organized as follows: for each Wikipedia article, we first model it as the $WST$ structure. Then we add all nodes of $WST$ except the root node to the set $S$. A node $d_i$ is randomly selected from $S$, and we define the depth of this node in $WST$ as $n$. Traversing from the root node to node $d_i$, all the titles on the path are put together to form a query $q^+$. The content of the node $d_i$ is defined as $D$. For the negative queries, we randomly select $n-1$ nodes from $S$, and concatenate the main title and subtitles of the selected nodes to define it as $q^-$. The relevance score is defined in Equation\ref{equ_score}. The loss function of the RWI task is defined as:

\begin{equation}
    \label{eq:RWI}
   \mathcal{L}_{RWI} = -\log_{}{    \frac{exp(score(q^+,D))}{exp(score(q^+,D))+\sum_{q\in S} exp(score(q,D))}} 
\end{equation}

\noindent where $q^+$ is the title, $D$ is the content of that article, and $S$ is the set of all negative queries generated from that article.

In this task, although both positive and negative queries contain the subtitles of the document, the positive query is more representative compared to the negative query. The model gives higher scores to the positive query through contrastive learning, so that the model can recognize the representative words in the text, and assign higher weights to these words if they are matched to the query. Therefore, through the RWI task, the model can learn how to identify the keywords in the text, which further leads to a better performance in the IR downstream task.
}
\subsection{Abstract Texts Identification (ATI)}
\label{s33}
In the ATI task, we utilize the abstract and the inner structure of Wikipedia. The abstract (the first section) of Wikipedia is regarded as the summarization of the whole article. Compared with other sections of the same article, the abstract is more likely to meet the user’s information needs when the query is the title. Therefore, we extract the title from the Wikipedia article as the query (denoted as $q$). Then the abstract of the same article is regarded as a positive document (denoted as $d^+$). For the negative ones, we use the other sections of the same article (denoted as $d^-$). The relevance score of a pseudo query-document pair is defined in Equation \ref{equ_score}. The loss function of the ATI task is defined as:


\begin{equation}
    \label{eq:ATI}
   \mathcal{L}_{ATI} = -\log_{}{    \frac{exp(score(q,d^+))}{exp(score(q,d^+)+\sum_{d\in S} exp(score(q,d))}} 
\end{equation}

\noindent where $q$ is the title of the article, $d^+$ is the abstract of the article, and $S$ is the set of all negative documents generated from that article.

\subsection{Long Texts Matching (LTM)}

After pre-training with RWI, ATI, and SRR tasks, Wikiformer acquires the ability to measure the relevance between a short text (query) and a long text. This can help the model better handle the vast majority of ad-hoc retrieval tasks. However, there are also scenarios involving "long queries", such as legal case retrieval and document-to-document search. In these scenarios, the model is required to match the relevance between two long texts. Fortunately, with the structured information of Wikipedia, especially hyperlinks, we can build a series of informative pseudo long query-document pairs. To be specific, we utilize the See Also section of Wikipedia which consists of hyperlinks that link to the other articles related to or comparable to this article. The See Also section is mainly written manually, based on the judgment and common sense of the authors and editors. Thus, we can obtain a series of reliable web pages that are highly related to the content of this page.

To this end, we design the Long Texts Matching (LTM) task to encourage the Wikiformer to learn the relevance matching ability between two long documents. Initially, we transformed the complete Wikipedia corpus into a graph structure by leveraging the interconnections provided by the 'See Also' links. This graph is designated as the See Also Graph (SAG). Each hyperlink in the See Also section can be formally represented as $(v_i, v_j)$, which means that $v_j$ appears in the See Also section of $v_i$. Consequently, $SAG$ can be defined as a \textit{directed graph}: $SAG = (V, E)$, where $E$ is the above-mentioned set of ordered pairs $(v_i, v_j)$ and $V$ is a set of Wikipedia articles. The order of an edge indicates the direction of hyperlinks. After building $SAG$, we use a contrastive sampling strategy based on the graph. For each node in $SAG$, we define its content as query $D$ and define all its adjacent nodes as positive documents $d^+$. We randomly select other documents as $d^-$. The relevance score of a pseudo query-document pair is defined in Equation \ref{equ_score}. The loss function of the LTM task is defined as:

\begin{equation}
    \label{eq:LTM}
   \mathcal{L}_{LTM} = -\log_{}{    \frac{exp(score(D,d^+))}{exp(score(D,d^+))+\sum_{d\in S} exp(score(D,d))}} 
\end{equation}

\noindent where $d^+$ is the adjacent articles, $D$ is the content of the original article, and $S$ is the set of all negative articles.

\subsection{Final Training Objective}
We add the loss of the proposed four tasks together as the overall loss of the model:

\begin{equation}
    \label{eq:final}
    \mathcal{L}_{final}=\mathcal{L}_{SRR}+\mathcal{L}_{RWI}+\mathcal{L}_{ATI}+\mathcal{L}_{LTM}
\end{equation}

\section{Experiments}

\subsection{Dataset Description}

For the {pre-training dataset}, we use the English Wikipedia (version 20220101). 
For the {downstream datasets}, we evaluate the performance of Wikiformer on five IR benchmarks. The basic statistics are shown in Table \ref{benchmark}. {MS MARCO Document Re-ranking}~\cite{nguyen2016ms} is a large-scale ad-hoc retrieval dataset with 0.37M queries and 3.2M documents. {TREC DL 2019}~\cite{craswell2020overview} shares the same document collection with MS MARCO but collects finer-grained human labels for 43 queries in the test set. {TREC Covid Round2}~\cite{roberts2021searching} is an ad-hoc retrieval dataset consisting of biomedical articles. It contains the May 1, 2020 version of the CORD-19~\cite{wang2020cord} document set and 35 queries written by biomedical professionals. {LeCaRD}~\cite{ma2021lecard} is a legal case retrieval dataset, consisting of 107 query cases and 10700 candidate cases. The queries in the LeCaRD dataset are the factual description part of a legal case, while the candidate documents are complete legal cases. {CAIL-LCR}\cite{CAIL} is a case retrieval dataset (document-to-document search) provided by CAIL 2022 consisting of 130 query cases and 100 candidate cases for each query case.

\begin{table}
\centering

  \caption{Basic statistics of our benchmark datasets}
  \label{benchmark}
{\tiny
  \begin{tabular}{ccccc}
    \toprule
    \textbf{Dataset} & \textbf{Genre} & \textbf{\#Queries} & \textbf{\#Documents} \\
    \midrule
    \textbf{MS MARCO} & web pages & 0.37M & 3.2M \\
   \textbf{ TREC DL 2019}  & web pages & 43 & 3.2M \\
    \textbf{TREC Covid} & biomedical & 35 & 59,851 \\
    \textbf{LeCaRD} & legal & 107 & 10,700 \\
    \textbf{CAIL-LCR} & legal & 130 & 13,000 \\
  \bottomrule 
\end{tabular}
}
\end{table}

\begin{table*}[]
\caption{The experimental results of Wikiformer and other baselines on three datasets in the zero-shot and fine-tuning setting. “*” denotes the result is significantly worse than Wikiformer with $ p < 0.05 $ level. The best results are in bold. “N” stands for nDCG. The zero-shot performance of both KNRM and Conv-KNRM methods is the same as randomized ranking. Therefore, their zero-shot performance is not shown in the table.}
\label{general_tab}
{\small
\begin{tabular}{cccccccccc}

\toprule

                   &            & \multicolumn{4}{c}{\textbf{Zero-shot}}                                                                            & \multicolumn{4}{c}{\textbf{Fine-tuned}}                                                                           \\
                   \midrule

                   &            & \multicolumn{2}{c}{\textbf{MS MARCO}} & \multicolumn{2}{c}{\textbf{TREC DL 2019}} & \multicolumn{2}{c}{\textbf{MS MARCO}} & \multicolumn{2}{c}{\textbf{TREC DL 2019}} \\
                   \midrule

Model Type         & Model Name & MRR@10                    & MRR@100                   & N@10                     & N@100                    & MRR@10                    & MRR@100                   & N@10                     & N@100                    \\
\midrule

Traditional Models & BM25       & 0.2656*                    & 0.2767*                    & 0.5315*                      & 0.4996*                      & 0.2656*                    & 0.2767*                    & 0.5315*                      & 0.4996*                      \\
                   & QL         & 0.2143*                    & 0.2268*                    & 0.5234*                      & 0.4983*                      & 0.2143*                    & 0.2268*                    & 0.5234*                      & 0.4983*                      \\
                   \midrule

Neural IR Models   & KNRM       & NA                        & NA                        & NA                          & NA                          & 0.1526*                    & 0.1685*                    & 0.3071*                      & 0.4591*                      \\
                   & Conv-KNRM  & NA                        & NA                        & NA                          & NA                          & 0.1554*                    & 0.1792*                    & 0.3112*                      & 0.4762*                      \\
\midrule

Pre-trained Models & BERT       & 0.1684*                    & 0.1811*                    & 0.3407**                      & 0.4316*                      & 0.3826*                    & 0.3881*                    & 0.6540                       & 0.5325*                      \\
                   & PROP\_WIKI & 0.2205*                    & 0.2321*                    & 0.4712*                      & 0.4709*                      & 0.3866*                    & 0.3922*                    & 0.6399*                      & 0.5311*                      \\
                   & PROP\_MS   & 0.2585*                    & 0.2696*                    & 0.5203*                      & 0.4810*                       & 0.3930*                     & 0.3980*                     & 0.6425*                      & 0.5318*                      \\
                   & Webformer  & 0.1664*                    & 0.1756*                    & 0.3758*                      & 0.4550*                       & 0.3984*                    & 0.4036*                    & 0.6479*                      & 0.5335                      \\
                   & HARP       & 0.2372*                    & 0.2465*                    & 0.5244*                      & 0.4721*                      & 0.3961*                    & 0.4012*                    & 0.6562                      & 0.5337                      \\
                   & ARES       & 0.2736*                    & 0.2851*                    & 0.5736*                      & 0.4752*                      & 0.3995*                    & 0.4041*                   & 0.6505*                      & 0.5353                      \\
\midrule

Our Approach       & Wikiformer & \textbf{0.2844}           & \textbf{0.2911}           & \textbf{0.5907}             & \textbf{0.5143}             & \textbf{0.4085}           & \textbf{0.4136}           & \textbf{0.6587}             & \textbf{0.5392}             \\

\toprule

\end{tabular}
}
\end{table*}

\subsection{Baselines for Comparison}
We consider three types of IR baselines for comparison, including traditional IR methods, Neural IR models, and pre-trained language models:

\noindent\textbf{Query Likelihood}~\cite{zhai2008statistical} is a language model based on Dirichlet smoothing. 

\noindent\textbf{BM25}~\cite{robertson2009probabilistic} is a highly effective retrieval model based on lexical matching.

\noindent\textbf{KNRM}\cite{xiong2017end} is an Interactive-based Neural Ranking Model that uses kernel-pooling to provide matching signals for each query-document pair.

\noindent\textbf{Conv-KNRM}\cite{dai2018convolutional} is a Convolutional Kernel-based Neural Ranking Model that fuses the contextual information of the surrounding words for relevance matching.

\noindent\textbf{BERT}~\cite{devlin2018bert} is a bi-directional Transformer based Pre-trained Language Model that has a powerful ability on contextual text representations.

\noindent\textbf{PROP\_MS}~\cite{ma2021prop} adopts the Representative Words Prediction (ROP) task to learn relevance matching from the pseudo query-document pairs. It is pre-trained on MS MARCO.\footnote{As PROP and B-PROP have similar performance, and B-PROP does not have a publicly available model checkpoint, therefore we only choose PROP as the baseline instead of selecting both of them.}

\noindent\textbf{PROP\_WIKI}~\cite{ma2021prop} adopts the same pre-training task as PROP\_MS. The only difference is that PROP\_WIKI is pre-trained on Wikipedia.

\noindent\textbf{HARP}~\cite{ma2021pre} utilizes the hyperlinks and anchor texts to generate pseudo query-document pairs and achieves state-of-the-art performance on ad-hoc retrieval.

\noindent\textbf{ARES} ~\cite{chen2022axiomatically} is a pre-trained language model with Axiomatic Regularization for ad hoc Search.

\noindent\textbf{Webformer} ~\cite{guo2022webformer} is a pre-trained language model based on large-scale web pages and their DOM (Document Object Model) tree structures.

\subsection{Implementation Details}
For the implementation of KNRM and Conv-KNRM, we use the OpenMatch\footnote{https://github.com/thunlp/OpenMatch} toolkit, and the 300d GloVe~\cite{pennington2014glove} vectors are used to initialize the word embeddings. For the implementation of BM25 and QL, we use the pyserini toolkit\footnote{https://github.com/castorini/pyserini}. For the hyperparameter of BM25, we set $k1 = 3.8$ and $b = 0.87$\footnote{This is the best hyperparameter we got after parameter searching.}. \textbf{Note that in our experiments, we use the scores of the BM25 and QL models to re-rank the candidate documents, rather than re-ranking the whole corpus.} For the implementation of BERT, we use the Pytorch version BERT-base released by Google\footnote{https://github.com/google-research/bert}. For the implementation of ARES, PROP\_MS, and PROP\_WIKI, we directly use the checkpoints released by the original paper. Since the original paper of Webformer and HARP did not release any checkpoints, we reproduce them on the same dataset based on their code and the details provided in their paper. 

To facilitate comparison with previous baselines, we adopted the same architecture as BERT-base. {This aligns with the settings of previous works such as ARES, HARP, PROP, B-PROP, and Webformer.} {To save computational resources during training, we initialized our model with BERT-base, following the same setting as previous works such as ARES and HARP. }We use the AdamW optimizer with a learning rate of 1e-5 in the first 50k steps and 5e-6 in the following steps. We set the warm-up ratio to 0.1. In the RWI, ATI, and SRR tasks, we set the maximum length of the query as 30 and the maximum length of the documents as 480. In the LTM task, we set the maximum length of both documents as 255. We trained our model on four Nvidia GeForce RTX 3090 GPUs for 60 hours. After training for 50k steps, we save the checkpoint every 5k steps and evaluate the zero-shot performance of each checkpoint on a subset of the MS MARCO training set which has no overlap with our test set. We select the best zero-shot performance checkpoint as the final model.

\subsection{Evaluation Methodology}For the two large-scale datasets MS MARCO and TREC DL 2019, we use Mean Reciprocal Rank at 10 and 100 (MRR@10 and MRR@100) for MS MARCO and normalized discounted cumulative gain at 10 and 100 (nDCG@10 and nDCG@100) for TREC DL 2019 as the evaluation metrics. For TREC Covid, we follow the setting of OpenMatch which re-ranks the top 60 candidates provided by the BM25-fusion method. We use precision at rank 5 (P@50) and nDCG@10 as the evaluation metrics for TREC Covid. For LeCaRD and CAIL-LCR datasets, we re-rank the candidate documents provided by the original dataset and use nDCG@5 and nDCG@15 as the evaluation metrics.

{For the significance test, we adopt Fisher’s randomization test \cite{fisher1936design,cohen1995empirical,box1978statistics} which is recommended for IR evaluation by previous work \cite{smucker2007comparison}.}

\subsection{Experimental Results}

\subsubsection{Zero-shot Performance}

Zero-shot performance is the performance of the model without any supervised data for fine-tuning. Thus, it directly reflects the effectiveness of the pre-training tasks. The experimental results are shown in Table \ref{general_tab}. We can see that Wikiformer outperforms all baselines on all evaluation metrics which shows the superiority of Wikiformer in the zero-shot setting. Based on the results, we also have the following findings:

Pre-trained models tailored for IR such as PROP, ARES, and Wikiformer perform significantly better than BERT in zero-shot settings. This shows the effectiveness of the pre-training tasks tailored for IR and that these models have indeed learned useful knowledge for relevance matching. Wikiformer performs the best among all the baselines in both benchmarks in zero-shot settings. Since the model architecture and parameter size of Wikiformer are the same as the other pre-trained models, this shows the effectiveness of our pre-training method. Besides, Wikiformer, Webformer, and PROP-Wiki are all pre-trained on the Wikipedia corpus. The superior performance of Wikiformer shows that it has made better use of Wikipedia and learned the rich knowledge that is helpful to solve IR problems through structured information on Wikipedia.

\subsubsection{Fine-tuned Performance}
  
Table\ref{general_tab} reports the performance of Wikiformer and other baselines after fine-tuning. Through the experimental results, we have the following findings:
(1) Although the performance of most pre-trained language models (PLMs) is inferior to traditional methods like BM25 and QL in the zero-shot setting, they surpass BM25 and QL significantly after fine-tuning. However, even after fine-tuning, Neural IR Models still underperform BM25 and QL. (2) On the MS MARCO dataset, IR PLMs consistently outperform BERT under the fine-tuning setting. This indicates that the knowledge acquired by IR PLMs during the pre-training stage remains valuable even after fine-tuning. HARP and Webformer, due to the incorporation of external knowledge such as hyperlinks, DOM Tree, and HTML tags, exhibit better performance than PROP-WIKI and PROP-MS. (3) Wikiformer significantly outperforms other baselines on both datasets. Note that the model structure and fine-tuning dataset for Wikiformer are the same as other baselines. Therefore, these experimental results indicate that Wikiformer has acquired more information retrieval knowledge during the pre-training stage compared to other baselines. This demonstrates the value of our pre-training task.

\subsubsection{Performance on Vertical Domains}

We conducted experiments on the legal domain dataset LeCaRD and CAIL-SCR as well as the biomedical domain dataset TREC Covid to explore the performance of Wikiformer in vertical domains. The experimental results are presented in Tables\ref{tab:lecard} and Table\ref{tab:covid}. The experimental results indicate that Wikiformer outperforms previous pre-trained models significantly in both the legal and biomedical domains. This suggests that Wikiformer possesses a domain-specific adaptability and effectiveness that allows it to excel in information retrieval tasks within these specialized fields. Its superior highlights the potential of utilizing Wikiformer for improving search and retrieval tasks across diverse domains.

\begin{table}[]
\centering
\caption{The experimental results of Wikiformer and other baselines on the TREC Covid rnd2 dataset. The best results are in bold.  “*” denotes the result is significantly worse than Wikiformer with $ p < 0.05 $ level. “N” stands for nDCG.}
\label{tab:covid}
\begin{tabular}{lcc}
\toprule
           & \multicolumn{2}{c}{\textbf{TREC Covid rnd2}} \\
           \midrule
           & \textbf{ Zero-shot N@10        }         & \textbf{Fine-tuned N@10}        \\
           \midrule
\textbf{QL}         & 0.4683*         & 0.4683*        \\
\textbf{BM25}       & 0.4792*         & 0.4792*        \\
\midrule
\textbf{BERT}       & 0.4018*         & 0.5580*         \\
\textbf{PROP\_MS }  & 0.4994*         & 0.5944*        \\
\textbf{PROP\_WIKI       } & 0.4137*         & 0.6104*        \\
\textbf{Webformer}  & 0.3845*         & 0.6032*        \\
\textbf{HARP}  & 0.4027*         & 0.5832*        \\
\textbf{ARES}       & 0.4993*         & 0.5969*        \\
\midrule
\textbf{Wikiformer} & \textbf{0.5449}         & \textbf{0.6197}       \\
\toprule
\end{tabular}
\end{table}

\subsubsection{Long Text Matching Performance}

The performance of Wikiformer and other baselines on LeCaRD and CAIL-LCR are reported in Table \ref{tab:lecard}. LeCaRD and CAIL-LCR are Chinese legal case retrieval tasks that have relatively long queries and candidate documents. Thus, experiments on these datasets can evaluate the long text-matching performance of Wikiformer and the baselines. Since there is no Chinese-centric pre-trained model tailored for IR so far, we only use traditional methods and a Chinese version of the BERT model~\cite{cui2021pre} as baselines.

The experimental results show that Wikiformer achieves better performance than traditional statistic methods BM25 and QL but also pre-trained language model BERT in long text-matching tasks. These experimental results highlight the potential of Wikiformer in effectively evaluating long-text similarity and also underscore the effectiveness of the proposed Long Text Matching (LTM) task.

\begin{table}[t]
\centering
\caption{The experimental results of Wikiformer and other baselines on LeCaRD and CAIL-LCR in fine-tuning setting. The best results are in bold.  “*” denotes the result is significantly worse than Wikiformer with $ p < 0.05 $ level. N@5 and N@15 respectively represent nDCG@5 and nDCG@15.}

\label{tab:lecard}

{\small
\begin{tabular}{lcccccc}
\toprule
           & \multicolumn{2}{c}{\textbf{LeCaRD}} & \multicolumn{2}{c}{\textbf{CAIL-LCR}} \\ \midrule 
           & \textbf{N@5}  & \textbf{N@15} & \textbf{N@5}    & \textbf{N@15} \\
\midrule
\textbf{BM25}       & 0.6843*  & 0.7303*  & 0.7105*     & 0.7490*   \\
\textbf{QL}         & 0.6906*  & 0.7411*  & 0.7389*     & 0.7756*  \\
\textbf{BERT}       & 0.7553*   & 0.7966*  & 0.7993*     & 0.8085  \\
\midrule
\textbf{Wikiformer} & \textbf{0.7722}   & \textbf{0.8073}  & \textbf{0.8095}     & \textbf{0.8134}  \\
\toprule
\end{tabular}
}
\end{table}

\subsection{Impact of the Training Data Size}
\begin{figure}[t]
\centering
    \includegraphics[width=\columnwidth]{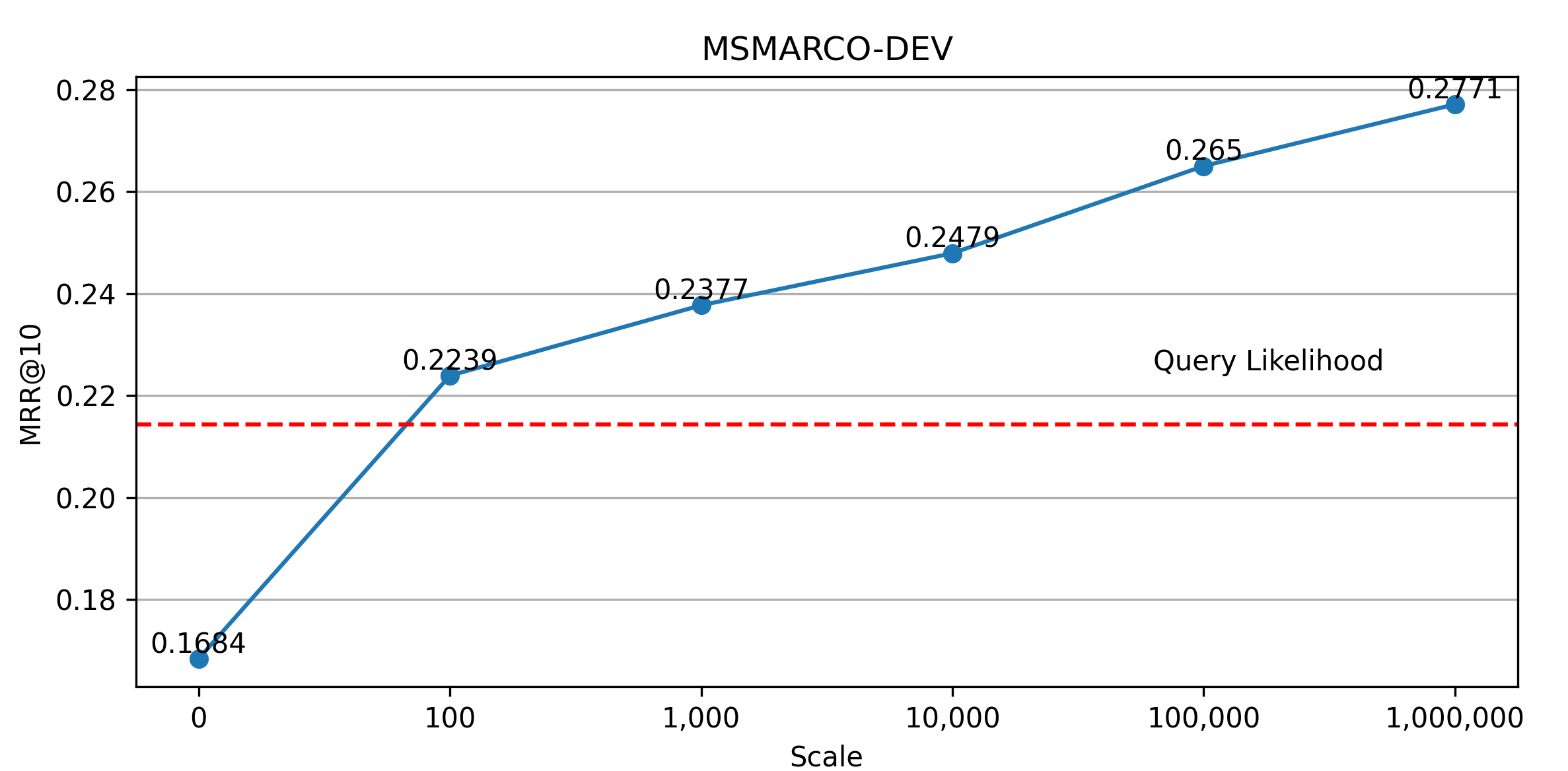}
    \caption{The performance of Wikiformer at different sizes of pre-training data sampled from the SRR task. The red dotted line shows the performance of Query Likelihood.}
    \label{scale}
\end{figure}

To investigate whether a larger training dataset enhances the performance of the pre-training phase, we evaluate the performance of Wikiformer on different sizes of training data varying from 100 to 1,000,000 pseudo query-document pairs. As shown in Figure \ref{scale}, Wikiformer surpasses the Query Likelihood model by pre-training with only 100 pseudo query-document pairs in the SRR task. This experimental result shows the effectiveness of our pre-training task and our proposed pseudo query-document pair sampling strategy. 

\begin{table}[]

    \caption{Ablation study results. The best results are in bold and the worst results are underlined.“*” denotes the performance is significantly better than the backbone model (BERT) with $ p < 0.01 $ level.}
    
    \label{table:ablation}

\centering
\small{
\begin{tabular}{lccc}
\toprule
                             & \multicolumn{2}{c}{\textbf{MS MARCO}} & \textbf{LeCaRD} \\
                             \toprule
                             & \textbf{MRR@10}   & \textbf{MRR@100}  & \textbf{nDCG@5} \\
                             \toprule
$w/o$ \textbf{SRR}                 & \underline{0.2334*}      & \underline{0.2441*}      & 0.7613*          \\
$w/o$ \textbf{RWI}                 & 0.2596*            & 0.2712*            & 0.7685*          \\
$w/o$ \textbf{ATI}                 & 0.2641*            & 0.2751*            & 0.7627*          \\
$w/o$ \textbf{LTM}                 & 0.2726*            & 0.2835*            & \underline{0.7574*}    \\
\midrule
\textbf{Before Pre-training} & 0.1684            & 0.1811            & 0.7553          \\
\textbf{All Four Tasks}      & \textbf{0.2844*}   & \textbf{0.2911*}   & \textbf{0.7722*} \\
\bottomrule
\end{tabular}
}
\end{table}

\subsection{Ablation Study}
To further analyze the effectiveness of each pre-training task, we conduct ablation experiments on MS MARCO (zero-shot) and LeCaRD (fine-tuned). The experimental results in table \ref{table:ablation} show that removing any pre-training tasks will lead to a drop in performance, indicating the effectiveness of each pre-training task on downstream IR tasks. On MS MARCO, among the four tasks, removing the SRR task leads to the largest performance degradation, which reveals that the hierarchical heading structure and the writing organization of Wikipedia contain valuable knowledge for ad-hoc retrieval which helps Wikiformer better at handling relevance matching. On LeCaRD, removing the LTM task leads to the largest performance degradation, which reveals that the LTM task is critical for improving the model's ability on long text-matching tasks.

\section{Conclusions}

In this paper, we propose Wikiformer, a pre-trained language model tailored for IR that achieves state-of-the-art performance. We propose several pseudo query-document pair sampling strategies based on the structured information on Wikipedia to leverage the wisdom of crowds brought by Wikipedia editors. Extensive experimental results and case studies verify the effectiveness of our pre-training methods. Results of the ablation study have also implied the effectiveness of all pre-training tasks. 

\section{Acknowledgements}

This work is supported by Quan Cheng Laboratory (Grant No. QCLZD202301), the Natural Science Foundation of China (Grant No. 62002194), and Huawei Poisson Lab.

\bibliography{aaai24}

\end{document}